\documentclass[preprint,proceedings]{rmaa}

\usepackage{paralist}

\usepackage{psfrag,color}

\newcommand{\rxj}{RX\,J0806}

\SetYear{2004}
\SetConfTitle{IAU Colloquium 194: Compact Binaries in the Galaxy and Beyond}

\title{On the Nature of the Binary Components of RX\,J0806.3+1527} 

\author{
  K. Reinsch,\altaffilmark{1,3} 
  V. Burwitz,\altaffilmark{2}
  and R. Schwarz\altaffilmark{1}}

\altaffiltext{1}{Universit\"ats-Sternwarte, G\"ottingen, Germany.}
\altaffiltext{2}{MPI f. extraterr. Physik, Garching, Germany.}
\altaffiltext{3}{Based on observations collected at the European Southern 
Observatory, Chile under proposal number 70.D-0683.} 

\shortauthor{Reinsch, Burwitz, \& Schwarz}
\shorttitle{Binary Components of RX\,J0806.3+1527}

\fulladdresses{
\item Vadim Burwitz: Max-Planck-Institut f\"ur extraterrestrische Physik, 
  P.O. Box 1312, D-85741 Garching, Germany (\email{burwitz@mpe.mpg.de}).
\item Klaus Reinsch and Robert Schwarz: Universit\"ats-Sternwarte, 
  Geismarlandstr. 11, D-37083 G\"ottingen, Germany
  (\email{reinsch,rsc@astro.physik.uni-goettingen.de}).
}

\listofauthors{K. Reinsch, V. Burwitz, \& R. Schwarz}
\indexauthor{Reinsch, K.}
\indexauthor{Burwitz, V.}
\indexauthor{Schwarz, R.}

\abstract{We present imaging circular polarimetry and near-infrared photometry 
of the suspected ultra-short period white-dwarf binary RX\,J0806.3+1527 
obtained with the ESO VLT and discuss the implications for a possible magnetic 
nature of the white dwarf accretor and the constraints derived for the nature 
of the donor star. Our $V$-filter data show marginally significant circular 
polarization with a modulation amplitude of $\approx 0.5$\% typical for 
cyclotron emission from an accretion column in a magnetic field of order 
10\,MG and not compatible with a direct-impact accretor model. The optical 
to near-infrared flux distribution is well described by a single blackbody with 
temperature $kT_{\rm bb} = 35000$\,K and excludes a main-sequence stellar donor 
unless the binary is located several scale heights above the galactic disk 
population.
}

\addkeyword{Polarization}
\addkeyword{Stars: binaries: close}
\addkeyword{Stars: individual: RXJ0806.3+1527}
\addkeyword{Stars: late-type}
\addkeyword{Stars: magnetic fields}

\begin{document}
\maketitle

\section{Introduction}
\label{sec:intro}

The soft X-ray discovered system RX\,J0806.3 +1527 (Beuermann et al. 1999)
has recently been suggested to be a semidetached white dwarf binary with 
a helium-degenerate secondary and the shortest known orbital period
Israel et al. (2002). If the 321 s pulse period (Israel et al. 1999,
Burwitz \& Reinsch 2001) is indeed the orbital period this system would lie 
close to the theoretical minimum period of white dwarf binaries and would 
be a suitable candidate for gravitational wave detection. 

Three flavors of the double-degenerate model (polar, direct accretor,
electric star) have been advanced to account for the observational 
characteristics (Cropper et al. 1998, Marsh \& Steeghs 2002, Wu et al. 2002). 
Alternatively, an interpretation of \rxj{} as a face-on, stream-fed 
intermediate polar has been advocated (Norton et al. 2003). 

We present the results of our recent imaging circular polarimetry and
near-infrared photometry and discuss the
implications for a possible magnetic nature of the white dwarf accretor
and the constraints derived for the nature of the donor star.

\section{Circular Polarimetry}
\label{sec:polarimetry}

Time-resolved circular polarimetry of \rxj{} has been obtained with FORS1 at 
the ESO VLT during 7 nights between December 2002 and February 2003. 
Our data cover a total of 3 h through a Bessel $V$ filter and 3 h in Bessel $I$
at $\approx 60$\,s time resolution corresponding to a phase resolution
$\Delta\Phi \approx 0.2$ of the 321\,s period.

\begin{figure}[!t]
  \includegraphics[width=\columnwidth,bb=33 148 580 497,clip=]{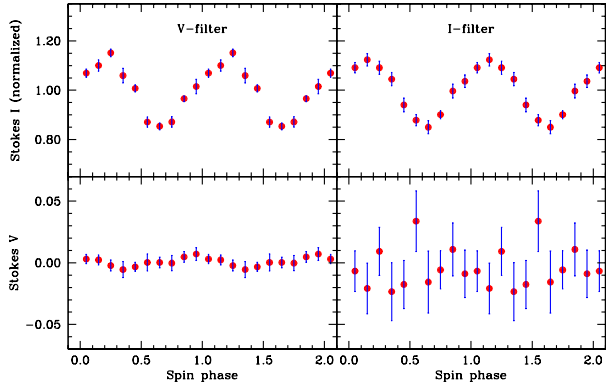}
  \caption{Normalized intensity variation ({\it upper panels\/}) and circular
  polarization ({\it lower panels\/}) of RX\,J0806.3+1527 observed in the 
  Bessel $V$ ({\it left side\/}) and Bessel $I$ ({\it right side\/}) filters, 
  respectively. All data have been phase folded with $P = 321.5393$\,s and 
  arbitrary zero point.}
  \label{fig:polarimetry}
\end{figure}

Stokes $V$ and $I$ data have been derived from the Bessel $V$ and $I$ images 
using aperture photometry (Fig. \ref{fig:polarimetry}). 
The 321\,s period phase-folded data show a marginally significant circular 
polarization with a modulation amplitude of $\approx 0.5$\% in Bessel $V$. 
The signal-to-noise of the $I$ filter data is much lower allowing us to derive
only an upper limit for the $I$-band circular polarization $V_{mean} \la 2$\%.

\section{Infrared Photometry}
\label{sec:IRphotometry}

2 hours of time-resolved jittered infrared imaging of \rxj{} in each of the 
$J$, $H$, and $Ks$ filter bands has been performed with ISAAC SW at the ESO 
VLT under photometric conditions during 5 nights between January and March 
2003. Effective integration times were 60 s per image. 

\begin{figure}[!t]
  \includegraphics[width=\columnwidth,bb=30 155 580 497,clip=]{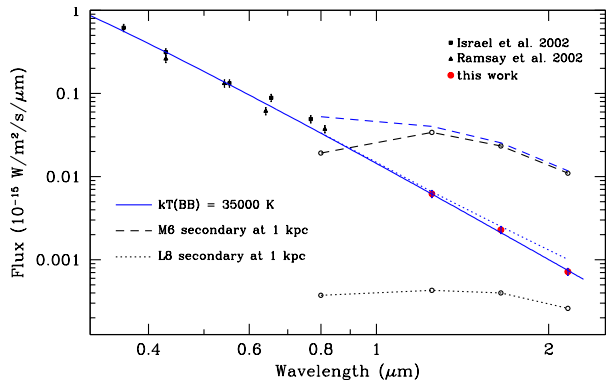}
  \caption{Optical and near-infrared flux distribution of RX\,J0806.3+1527. 
  The {\it straight line\/} represents a $kT_{\rm bb} = 35000$\,K blackbody. 
  For comparison the flux contributions and total fluxes expected for a
  late-type secondary at 1\,kpc distance are shown ({\it dashed line\/}: M6 
  star, {\it dotted line\/}: L8 star).}
  \label{fig:optIRflux}
\end{figure}

The average fluxes of \rxj{} in the $J$, $H$, and $Ks$ filter bands
significantly extend the optical flux distribution obtained by Israel 
et al. (2002) and Ramsay et al. (2002) (Fig. \ref{fig:optIRflux}). The optical
to near-infrared flux distribution is well described by a single blackbody with 
temperature $kT_{\rm bb} = 35000$\,K.

\section{Discussion}
\label{sec:discussion}

The circular polarization detected in the $V$-filter data is comparable to 
that seen in some intermediate polars (e.g. Buckley et al. 1995). This is 
characteristic of cyclotron emission originating from an accretion column in 
a magnetic field of order 10\,MG and would be compatible with all currently
discussed models except for the direct accretor.

The optical-to-infrared flux distribution excludes a main-sequence stellar 
donor unless the binary is located several scale heights above the galactic 
disk population. If the system is at a distance of 1\,kpc (i.e. at 
approximately twice the galactic scale height) the secondary must be of 
spectral type L8 or later that its flux contribution can be hidden in the 
observed spectrum. This is difficult to conceal in an intermediate polar 
interpretation of \rxj{} which implies a Roche-lobe filling stellar companion.

On the other hand, the question whether other periods are present in the
system (as expected for an intermediate polar) is not well settled and needs
further observations. Furthermore, the evidence for Hydrogen in the optical 
spectrum of \rxj{} implies severe problems for all ultra-short period binary 
models which require a Helium rich or a Helium degenerate secondary.

Concluding, none of the models proposed so far fits well with all available
observations and the true nature of RX\,J0806.3+1527 must still be considered 
open.

\acknowledgements

We thank ESO staff members at Paranal observatory for obtaining the data
as part of an ESO Service Mode run. This project has been supported in part
by BMBF/DLR grant 50\,OR\,0206.

\end{document}